\documentclass{aa}

\usepackage{graphicx}
\usepackage[varg]{txfonts}
\usepackage{natbib}

\usepackage[modulo, switch]{lineno}

\def\arcsec{\hbox{$^{\prime\prime}$}}

\newcommand{\sur}[1]{\ensuremath{^{\rm #1}}}
\newcommand{\sous}[1]{\ensuremath{_{\rm #1}}}

\newcommand{\etal}{et~al. }

\newcommand{\caiih}{Ca\,{\sc ii}\,{\sc H}}

\newcommand{\caii}{Ca\,{\sc ii}}
\newcommand{\caxvii}{Ca\,{\sc xvii}}

\newcommand{\sivii}{{Si~{\sc vii}\,}}
\newcommand{\six}{{Si~{\sc x}\,}}
\newcommand{\fei}{{Fe~{\sc i}\,}}

\newcommand{\feviii}{{Fe~{\sc viii}\,}}

\newcommand{\fex}{{Fe~{\sc x}\,}}
\newcommand{\fexi}{{Fe~{\sc xi}\,}}
\newcommand{\fexii}{{Fe~{\sc xii}\,}}
\newcommand{\fexiii}{{Fe~{\sc xiii}\,}}

\newcommand{\ov}{{O~{\sc v}\,}}

\newcommand{\mgv}{{Mg~{\sc v}\,}}
\newcommand{\mgvi}{{Mg~{\sc vi}\,}}
\newcommand{\mgi}{{Mg~{\sc i}\,}}
\newcommand{\mgvii}{{Mg~{\sc vii}\,}}

\newcommand{\heii}{{He~{\sc ii}\,}}

\newcommand{\nii}{{Ni~{\sc i}\,}}

\newcommand{\Ha}{${\rm H\alpha}$ }
\newcommand{\ha}{${\rm H\alpha}$ }

\begin{document}

\title{Emergence of small-scale magnetic flux in the quiet Sun}

       \author{I. Kontogiannis\inst{1}
        \and G. Tsiropoula\inst{2}
        \and K. Tziotziou\inst{2}
        \and C. Gontikakis\inst{3}
        \and C. Kuckein\inst{1}
        \and M. Verma\inst{1}
        \and C. Denker\inst{1}}

\institute{
Leibniz-Institut f\"ur Astrophysik Potsdam (AIP), An der Sternwarte 16, 14482, Potsdam, Germany \email{ikontogiannis@aip.de}\label{inst1} \and
Institute for Astronomy, Astrophysics, Space Applications and Remote Sensing, National Observatory of Athens, 15236 Penteli, Greece\label{inst2} \and
Research Center for Astronomy and Applied Mathematics (RCAAM), Academy of Athens, 4 Soranou Efesiou Street, 11527 Athens, Greece\label{inst3}
}

\date{Accepted 29/11/2019}

\abstract{We study the evolution of a small-scale emerging flux region (EFR) in the quiet Sun, from its emergence in the photosphere to its appearance in the corona and its decay.}
{We track processes and phenomena that take place across all atmospheric layers; we explore their interrelations and compare our findings with those from recent numerical modelling studies.}
{We used imaging as well as spectral and spectropolarimetric observations from a suite of space-borne and ground-based instruments.}
 {The EFR appears in the quiet Sun next to the chromospheric network and shows all morphological characteristics predicted by numerical simulations. The total magnetic flux of the region exhibits distinct evolutionary phases, namely an initial subtle increase, a fast increase with a cotemporal fast expansion of the region area, a more gradual increase, and a slow decay. During the initial stages, fine-scale G-band and \caiih\ bright points coalesce, forming clusters of positive- and negative-polarity in a largely bipolar configuration. During the fast expansion, flux tubes make their way to the chromosphere, pushing aside the ambient magnetic field and producing pressure-driven absorption fronts that are visible as blueshifted chromospheric features. The connectivity of the quiet-Sun network gradually changes and part of the existing network forms new connections with the newly emerged bipole. A few minutes after the bipole has reached its maximum magnetic flux, the bipole brightens in soft X-rays forming a coronal bright point. The coronal emission exhibits episodic brightenings on top of a long smooth increase. These coronal brightenings are also associated with surge-like chromospheric features visible in H$\alpha$, which can be attributed to reconnection with adjacent small-scale magnetic fields and the ambient quiet-Sun magnetic field.}
{The emergence of magnetic flux even at the smallest scales can be the driver of a series of energetic phenomena visible at various atmospheric heights and temperature regimes. Multi-wavelength observations reveal a wealth of mechanisms which produce diverse observable effects during the different evolutionary stages of these small-scale structures. }


\authorrunning{Kontogiannis \etal}
\titlerunning{Flux emergence in quiet Sun}

\maketitle

\section{Introduction}

The term ``flux emergence'' is used to denote the appearance of new magnetic field concentrations at the solar surface. This process introduces new magnetic field into the solar atmosphere and takes place across a range of spatial and temporal scales. At the largest scales, flux emergence results in the formation of complex active regions \citep{vanDriel_Gesztelyi15}, which may last for days or even weeks, while at supergranular scales it leads to smaller ``ephemeral'' (active) regions \citep{harvey73,Hagenaar01} with lifetimes of a few hours or days. At even smaller scales, flux emergence manifests itself as granular-scale events \citep{depontieu02} that form low-lying magnetic loops at the internetwork \citep{martinez_gonzalez07} or produce energetic events within active regions \citep[see e.g.,][and references therein]{toriumi17}.

The magnetic field configuration that is produced through emergence processes partakes in intricate interactions across all atmospheric layers either with the ambient magnetic fields and the plasma or with itself. Furthermore, the magnetic field can channel, store, and release vast amounts of energy, powering a plethora of solar phenomena. The study of emergence events through observations and simulations is therefore paramount to understanding the dynamical coupling between layers and observed structures in the solar atmosphere. During the past two decades, with the advent of  instruments of both high resolution and precision, and the development of sophisticated, realistic simulation codes, our understanding of the processes that take place during these events has increased significantly \citep{archontis12,cheung_isobe14,schmieder14}. 

\begin{figure}[htp]
\centering
  \includegraphics[width=9.5cm]{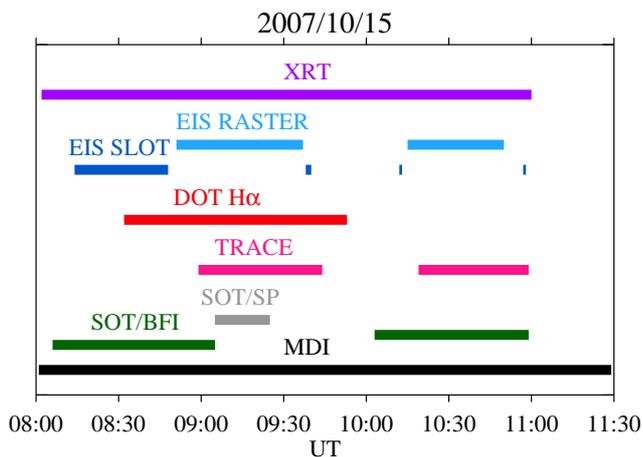}
  \caption[]
    {Temporal coverage of the solar disk center and the ROI by each instrument of the coordinated campaign on 15 October, 2007.}
    {\label{fig:obs}}
\end{figure}

Simulations, which include radiative transfer processes and thermodynamics effects, have offered a realistic overview of flux emergence from the convection zone to the corona \citep{cheung07,cheung08,martinez_sykora08,martinez_sykora09,tortosa_andreu09,moreno_insertis18}.
Twisted, coherent flux tubes can rise bodily from the interior to the photosphere because of buoyancy instabilities. Their interaction with the subphotospheric flow field may result in a considerable distortion of the flux tubes. Thus, depending on their initial magnetic strength, they may emerge coherently or in fragments through a series of small-scale events, producing a pattern of undulating magnetic field lines at the photosphere of emerging flux regions (EFR). The smallest of these events do not seem to affect the granulation pattern but are passively advected by photospheric flows. However, granulation and reversed granulation are affected considerably by stronger magnetic fields: granules are significantly elongated along the flux tube axes, as the latter pass through the surface (abnormal granulation), whereas, upon meeting the lower chromosphere, dark elongated patches appear over the abnormal granules with brightenings often appearing at their periphery \citep{tortosa_andreu09}. As the emerging magnetic flux continues to rise, it pushes the ambient magnetized layers of the upper atmosphere, carrying bubbles of colder chromospheric plasma upwards \citep{martinez_sykora08}. The interaction between the newly emerged and the ambient magnetic fields leads to Joule heating along the magnetic field lines \citep{martinez_sykora09}. 

Observations have largely corroborated modeling predictions regarding the impact on granulation and reversed granulation, lower chromospheric brightenings, and the recurrent character of small-scale flux emergence \citep{cheung08,guglielmino08,palacios12}. Spectropolarimetric measurements also confirm the rise of $\Omega$-loops through the photosphere. When observed at the disk center, the typical spectropolarimetric imprint of these events in the photosphere consists of an initial linear polarization patch (horizontal magnetic field) which is followed by flanking patches of opposite-polarity circular polarization (vertical magnetic flux). These opposite-polarity patches then separate while the linear polarization signal gradually fades \citep{centeno07,gomory10,guglielmino12}. \citet{martinez_gonzalez09} followed the evolution of such small-scale events down to two orders of magnitude weaker than ephemeral regions, that is, about $10^{16}$ \,Mx compared to about $10^{18}$ \,Mx, which is the lower limit of magnetic fluxes for ephemeral regions \citep{zwaan87}. Roughly a quarter of these flux tubes clearly appeared in the strong chromospheric lines of \mgi\,$b{2}$  and \caiih\, and it is questionable whether the smallest of them would appear at the upper chromosphere. 

However, small-scale magnetic flux emergence affects a range of heights and temperature regimes and multi-wavelength studies are therefore necessary. Such studies revealed that the interaction with the ambient magnetic field, with other pre-existing or emerging magnetic fields, or between field lines of the same structure being submerged by the convective flows leads to various energetic events in different heights and temperatures. \citet{guglielmino10} followed the emergence of a small-scale magnetic flux within an active region with high-resolution observations spanning from the photosphere to the corona along with measurements of the photospheric magnetic field. Throughout a two-hour period, they observed a series of EUV and X-ray brightenings and \ha surges, which were attributed to reconnection at the low chromosphere between the emerging region and a parasitic polarity. High-resolution spectral imaging and spectropolarimetry in the \caii\,8542\,{\AA} line revealed the existence of magnetized bubbles in small-scale flux emergence within active regions \citep{ortiz14,delacruzrodriguez15}. As these bubbles rise and interact with the ambient magnetic field, they exhibit signatures of heating and can be associated with explosive events in the chromosphere and the transition region \citep{ortiz16}. In the same context, \citet{centeno17} studied the interaction of two small-scale flux emergence events within an active region and found UV brightenings at the interaction region, which they attributed to magnetic reconnection.

This kind of interaction between consecutive small-scale flux emergence events is very common in active regions during the reconfiguration of the overall magnetic field of the region and leads to energetic events such as surges, Ellerman's bombs \citep[EBs;][]{ellerman17,georgoulis02}, and UV bursts \citep{peter14,young18}. The connection between these phenomena remains a topic of interest \citep[see e.g., ][]{vissers15}, because reconnection at different heights (and hence seen in different observables) produces different amounts of plasma heating, which may be observed in different locations within the same region. This is illustrated in the most recent realistic simulations, which show that the interaction between the newly emerged and the ambient magnetic fields can produce jet-like events during different phases of flux emergence \citep{mactaggart15}, with observable effects in different temperature regimes, in agreement with observations \citep{nobrega16,nobrega17,guglielmino18,guglielmino19}.   

In \citet{kontogiannis18} we used a dataset of multi-wavelength observations to explore the structure and dynamics of a quiet solar region at the disk center. In this work we use the same dataset to focus on the evolution of an EFR at mesogranular scale within the same region. Located at the center of the solar disk away from dominant large-scale, overlying magnetic structures, it allows us, for the first time, to follow the overall morphology and evolution of emerging magnetic flux from the photosphere up to the corona in a quiescent magnetic environment. We use high-resolution photospheric spectropolarimetry and EUV spectroscopy to infer the atmospheric properties of the EFR and spectral imaging to monitor its dynamics at the chromosphere. The work is organized as follows: Section 2 provides an overview of the observations and their analysis, and Section~3 describes the evolution of flux emergence as the new region successively enters the photosphere, the chromosphere, and the corona. We discuss our findings in Section 4 and put them in the context of the current understanding of flux emergence.

\section{Observations and analysis}

The observations used in this study were obtained between 08:00 and 11:00\,UT on October 15, 2007, during a coordinated observing campaign which included space-borne and ground-based instruments. The target of the campaign was a quiet-Sun region at the solar disk center. Figure~\ref{fig:obs} summarizes the temporal coverage of the observed region by each instrument while Table~\ref{Table:t1} contains the characteristics of the corresponding datasets. 

All instruments of the Hinode mission were involved in the campaign. The Solar Optical Telescope \citep[SOT,][]{sot} provided time-series of broadband \caiih\ and G-band images with an image scale of 0.054\arcsec pixel$^{-1}$, and a cadence of 110\,s. 

\begin{table}
\caption{Instruments used in this study.}
\setlength{\tabcolsep}{4.pt}
\begin{tabular}{lccc}
\hline
           & Bandpass           &         & Image                 \\
Instrument & Spectral area      & Cadence & Scale                   \\
           &  [{\AA}]           & [s]     & [arcsec]                \\
\hline
MDI        &  \nii 6767.8        & 60      &           0.6           \\
SOT/BFI    & \caiih\ / G\,band    & 110     &         0.054           \\ 
SOT/SP     & \fei 6301.5 \& 6302.5          &(rasters)& 0.32         \\              
EIS        &    EUV 40\arcsec slots        & 65      &   1          \\
           &     + 2 rasters      & -- &     2$\times$1             \\
XRT        &   ``C-Poly''         & 25      & 1                         \\
TRACE      &    1550, 1600, 1700  & 30      & 0.5                       \\
DOT        &H$\alpha$ line center, $\pm$0.35, $\pm$0.7 &30  & 0.109  \\
\hline
\end{tabular}
\label{Table:t1}
\end{table}

The properly aligned images of G-band and \caiih\ served as input for the computation of horizontal proper motions. This was done using the local correlation tracking (LCT) algorithm of \citet{Verma2011} and \citet{Verma2013}, which is based on ideas put forward by \citet{November1988}. Before the application of LCT a few pre-processing steps were carried out, such as subpixel alignment of images and removal of the signature of the five-minute oscillations by applying a 3D Fourier filter. The LCT velocities were computed over image tiles of $64 \times 64$ pixels with a Gaussian kernel that has a full-width-at-half-maximum (FWHM) of 1200\,km. Consecutive images were used to calculate LCT maps:  the cadence was $\Delta t=110$~s, and the computed maps were then averaged over a time period of $\Delta T= 62$~min. We chose this value of $\Delta T$ for averaging the LCT maps to bring out the persistent motions in this small-scale EFR. Although the cadence of the data was slightly higher than the 60--90\,s recommended by \citet{Verma2011}, it was still sufficient to reveal flow details in the high-resolution Hinode images.

The SOT Spectropolarimeter \citep[SP,][]{sot_sp} was operating in ``normal mode'' and performed two scans of the region at 09:05\,UT and 09:15\,UT. This provided Stokes spectra of the photospheric \fei\ 6301 and 6302\,{\AA} lines, with an image scale equal to 0.32\arcsec. The HAO/CSAC team\footnote{\url{http://www.csac.hao.ucar.edu/csac/dataHostSearch.jsp}} provided the level-1 data which were then used as input for the spectropolarimetric inversions.

We inverted both \fei\ lines simultaneously using the Stokes Inversions based on Response functions (SIR) code \citep{sir}. The code derives physical parameters such as temperature, magnetic field strength, inclination, azimuth, and LOS velocity as a function of continuum optical depth. Normally, the overall morphology of this quiet Sun region can be retrieved accurately with a simple inversion scheme (e.g., two nodes in temperature, magnetic field strength and inclination). However, in order to fit some irregular Stokes profiles found in our FOV, a more complex approach was required. Therefore, for the inversions we followed a similar strategy as described in \citet{ruizcobo13}. We used up to five nodes (optical depth points) in temperature, magnetic field strength, LOS velocity, and inclination, and two nodes for the azimuth. In addition, the spectral PSF and a stray light profile, which was computed as the average intensity profile across the whole FOV, were included during the inversion process. For this average profile a Gaussian fit was performed to obtain the center of the line. Since the Gaussian fit takes into account the whole spectral line, we expect that the retrieved center represents all optical depths. This central wavelength was then subtracted from the wavelength array and marks our zero velocity for all optical depths.
To avoid local minima, we initialized the code with four different atmospheric models, namely HSRA \citep{gingerich71}, FAL-C, FAL-E, and FAL-F \citep{fontenla06}. The first two represent quiet Sun, while the latter two represent network lane and active network, respectively.
We kept the best result according to a
$\chi^2$-test between the observed and synthetic Stokes profiles. 

The Extreme Ultraviolet Imager Spectrometer \citep[EIS;][]{eis} provided time-series of 40\arcsec\ slots as well as two raster scans taken with the 2\arcsec\-slit. We performed a differential emission measure (DEM) analysis using the spectral lines observed with EIS (Table~\ref{Table:t2}).
The strong \heii 256.32\,\AA\ spectral line was not used in the DEM analysis for two main reasons. First the helium line is blended with \six\ 256.37\,\AA, \fexiii 256.42\,\AA, and \fexii 256.41\,\AA, \citep{young07}, and their elimination is difficult; although not impossible in quiet-Sun conditions. Most importantly, the \heii lines show an important discrepancy between observed and calculated intensities \citep{giunta15}, which is a problem that we cannot overcome.
For the treatment of the complex spectral structure in the 192.8\,--\,193\,\AA\ range, following \citet{levens15}, we assumed that the \caxvii\ 192.85\,\AA\ has negligible emission in a quiet-Sun area. In the fit we included the three \ov\ lines at 192.75\,\AA, 192.80\,\AA,\ and 192.90\,\AA\, whose relative intensities and wavelengths were constrained according to \citet{young07}, and the \fexi lines at 192.63, 192.80\,\AA. 

The X-Ray Telescope \citep[XRT;][]{xrt} provided time-series of soft X-ray filtergrams of the solar disk center, taken with the ``C-poly'' filter with 25\,s cadence and 1$\arcsec$ spatial scale. During the observing campaign, XRT was interrupting the regular observations to obtain full-disk filtergrams, which resulted in a nonregular acquisition time. These full-disk filtergrams were removed and were not considered in the present analysis.

\begin{table}
\caption{EIS spectral lines. Those marked ``DEM'' were used for the DEM analysis}
\begin{tabular}{llll}
Ion & Wavelength ({\AA}) & log(T)& comment \\ \hline
\heii & 256.32 & 4.7    & blended \\
\ov  & 192.75  &  5.25   &   \\
\ov  & 192.80  &  5.25   &    \\
\ov  & 192.90  &  5.25   & DEM  \\
\mgv  & 276.57  & 5.45    & weak line DEM\\
\mgvi & 269.00  & 5.65    & DEM\\
\feviii & 185.21 & 5.65   & DEM \\
\mgvii & 278.39  & 5.8    & DEM\\
\sivii & 275.35  & 5.8   & DEM\\
\fex  & 184.54   & 6.05   & DEM\\
\fexi & 192.63   & 6.15   & DEM\\
\fexi & 192.80   & 6.15   & DEM\\
\fexii & 195.12  & 6.2   & DEM\\ \hline
\end{tabular}
\label{Table:t2}
\end{table}

The Michelson Doppler Imager \citep[MDI;][]{mdi} onboard the Solar and Heliospheric Observatory (SoHO) was observing the solar disk center throughout the entire duration of the coordinated observations. The MDI was operating in high-resolution mode, producing context line-of-sight (LOS) magnetograms with an image scale of 0.6\arcsec\ and a cadence equal to 60\,s. These data allow us to follow the evolution of the photospheric LOS component of the magnetic field, $B\sous{LOS}$, in the region-of-interest (ROI). 

The Transition Region and Coronal Explorer \citep[TRACE][]{trace98,trace99} provided imaging of the same region in the 1550, 1600, and 1700\,{\AA} channels, recording emission mostly coming from a region around the temperature minimum in quiet-Sun conditions. The image scale of TRACE is 0.5\arcsec\ pixel$^{-1}$ while the cadence is 30\,s. During the observing sequence, TRACE also acquired seven images in the 171\,{\AA}, only the last of which shows the response of the lower corona to the event. Therefore, their further use would not add any extra information to the analysis. TRACE observations were useful during the alignment process but were not further utilized in the analysis.  

\begin{figure*}[htp]
\centering
  \includegraphics[width=18cm]{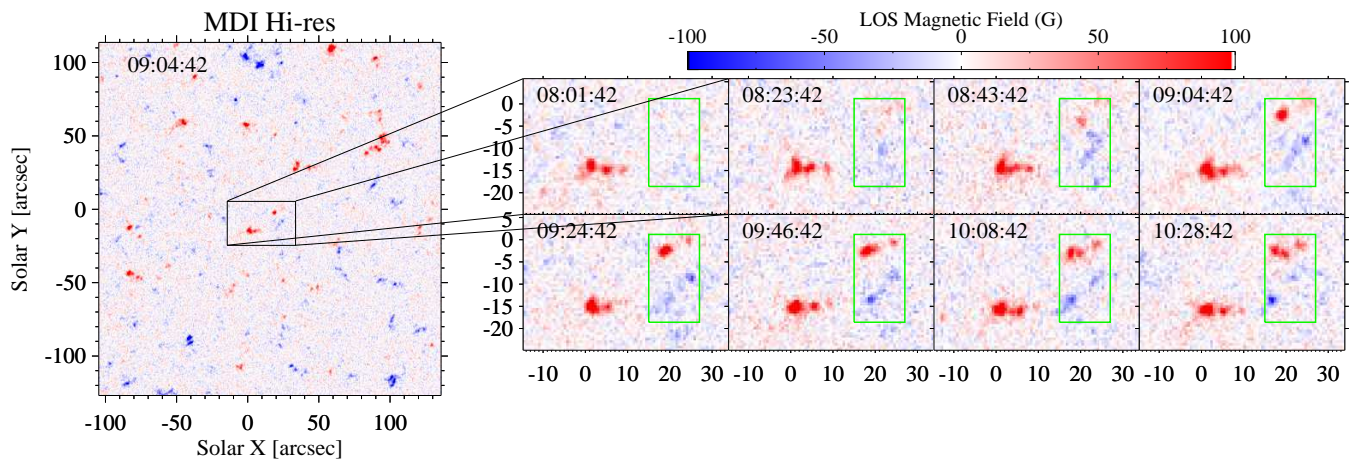}
  \caption[]
    {Left: Context magnetogram at solar disk center taken by the MDI. The longitudinal magnetic field strength is scaled between $\pm$100\,G to enhance the visibility of quiet-Sun features.  Right: Snapshots of the ROI hosting a well-formed network boundary and the EFR (marked by the green rectangle). All maps are aligned with solar north.}
    {\label{fig:mdi_snaps}}
\end{figure*}

\begin{figure}
\centering
  \includegraphics[width=9cm]{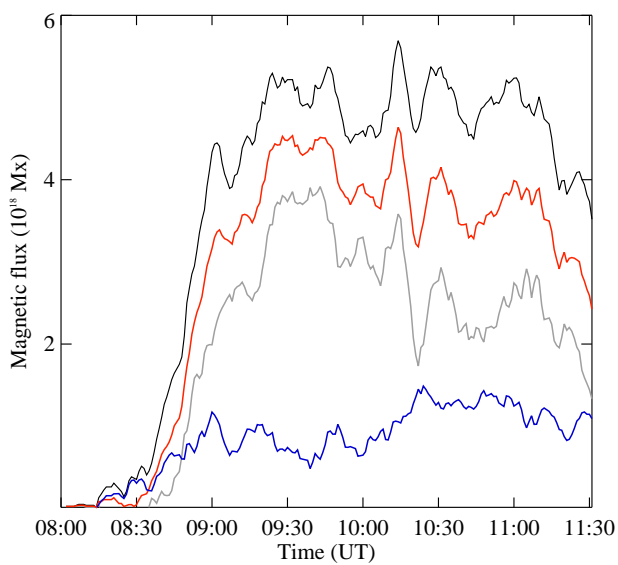}
  \caption[]
    {Temporal evolution of the total unsigned and net magnetic flux (solid and gray lines respectively) as well as the total positive and negative (red and blue lines respectively), as measured by the MDI, within the region marked by the green rectangle in Fig.~\ref{fig:mdi_snaps}. All curves are smoothed by a seven-minute running average.}
    {\label{fig:mag_evo}}
\end{figure}

Chromospheric spectral imaging observations were taken by the ground-based Dutch Open Telescope \citep[DOT;][]{rutten04a} located in LaPalma, Spain. The DOT produced time-series of speckle-reconstructed images in five wavelength positions along the \Ha profile, namely $\pm$0.70\,{\AA}, $\pm$0.35\,{\AA,} and line center. The image scale of the \ha observations is 0.109\arcsec pixel$^{-1}$ and the cadence is 30\,sec. The intensities at $\pm0.7$\,\AA\ were used to calculate maps of the Doppler signal (DS), according to the equation:

\begin{equation}
DS=\frac{I(+0.7)-I(-0.7)}{I(+0.7)+I(-0.7)}.
\label{equation:ds}
\end{equation}

\noindent The DS is a parametric description of the chromospheric LOS velocity field \citep{tsirop00}, and in this formulation positive (negative) values denote upward (downward) motions. 

Standard SolarSoft \citep{bentley98,freeland98} IDL routines were used to reduce the data from the space-borne instruments and perform flat-field and dark corrections, spike removal, and alignment of the data cubes. All observations were then co-aligned to a common temporal and spatial reference. The ground-based observations were rotated by $26^\circ$ to match solar north. The bright points of the network were then used as a reference to align the images. These are clearly discernible in the blue wing of H$\alpha$, in \caiih,\ and the TRACE channels, making co-alignment between these datasets relatively straightforward. The MDI magnetograms were co-aligned with the TRACE filtergrams, which have a similar image scale. The EIS observations are then co-aligned by comparing the \heii\ 40\arcsec\ slot images with the \caiih\ and TRACE filtergrams. Given the known shift between the two wavelength ranges of EIS, which is about 16\arcsec\ in the N--S and about 2\arcsec\ in the E--W direction\footnote{see e.g., \url{http://solarb.mssl.ucl.ac.uk:8080/eiswiki/}}, one can then co-align the short-wavelength range with the rest of the dataset. The \fexii slot images are co-aligned with the XRT images since they represent roughly the same coronal temperatures and structures. Further details on the observations, the speckle reconstruction procedure of the \ha observations, and some preliminary reduction steps are given in \citet{kont10b,kontogiannis11,kontogiannis18}.

\section{Results}

\subsection{Magnetic field evolution}
\label{sec:mdi}

A context magnetogram of the solar disk center taken at 09:04\,UT on October 15 is shown in Fig.~\ref{fig:mdi_snaps}, where the ROI is marked with a black rectangle. A positive-polarity cluster of the network is already formed at about (0,-15) and persists throughout the entire duration of the coordinated observations. Under the influence of the photospheric flow field, this network and all magnetic concentrations of the ROI evolve and interact. To the right of this positive-polarity magnetic cluster, a small-scale magnetic region started to emerge (green rectangle in Fig.~\ref{fig:mdi_snaps}). 

Initially, the EFR appeared as a small bipole (08:23\,UT), which then expanded and evolved into a well-defined compact, positive-polarity patch and a series of smaller scattered negative-polarity patches at the lower left part of the region (e.g., 08:43--09:04\,UT). These negative-polarity magnetic points gradually approached the pre-existing network region (09:24\,UT). One hour after its first appearance, a second positive-polarity patch appeared at the upper right of the region (09:24\,UT). Towards the end of the observations, the positive polarity started to fragment and the resulting smaller parts approached the remaining negative polarity (10:28\,UT).

\begin{figure*}[htp]
\centering
  \includegraphics[width=18cm]{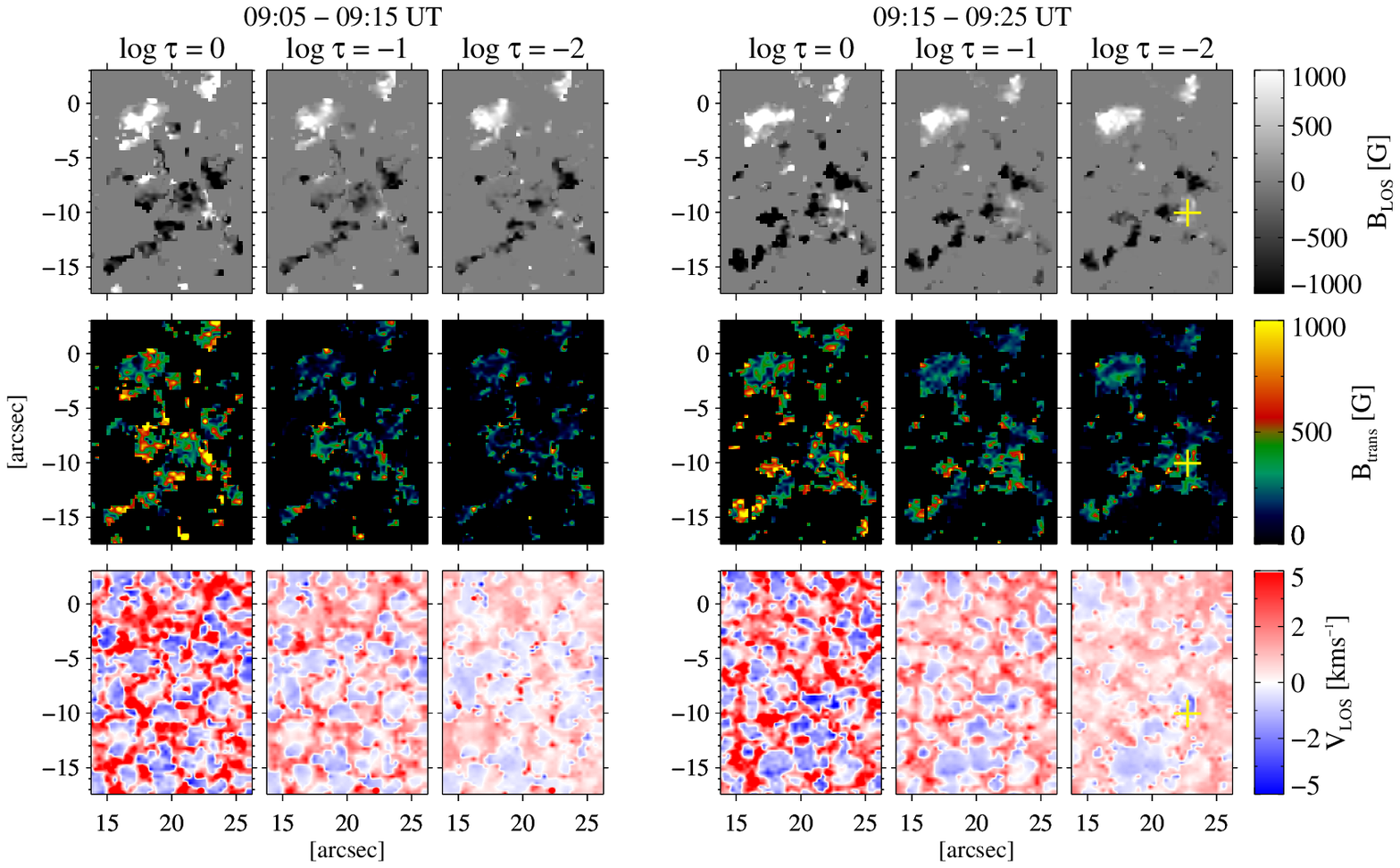}
  \caption[]
    {Results of the spectropolarimetric inversions with SIR, for the raster scans of Hinode/SP taken at 09:05\,UT (left) and 09:15\,UT (right). Columns correspond to different $\tau$ values while rows from top to bottom correspond to the LOS and transversal component of the magnetic field, $B\sous{LOS}$ and $B\sous{trans}$ (which correspond to the vertical and horizontal components, correspondingly, since the region is located at the disk center), and the LOS component of the Doppler velocity, $V\sous{LOS}$. The yellow crosses mark the pixel whose profiles are shown in Fig.~\ref{fig:profs}. Magnetic field results are shown for pixels with total polarization higher than 5\,\%. The magnetic field values in the map are clipped between $\pm$1000\,G and the values of the velocity between $\pm$5\,km/s. }
    {\label{fig:spmag}}
\end{figure*}

\begin{figure}[htp]
\centering
  \includegraphics[width=9cm]{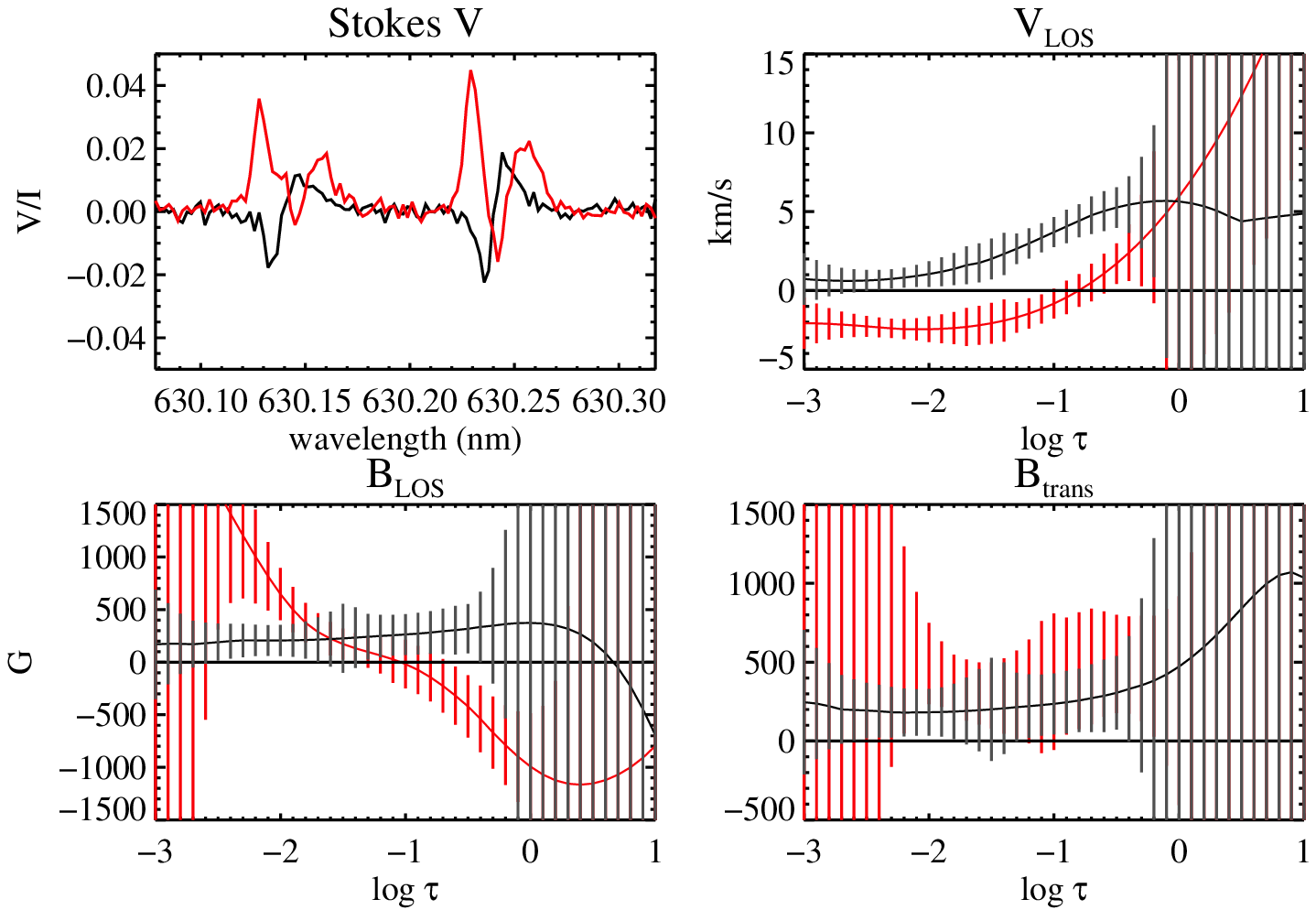}
  \caption[]
    {From upper left to lower right: Stokes \textit{V} profiles of the \fei\,6301\,{\AA} and 6302\,{\AA} lines, $V\sous{LOS}$, $B\sous{LOS}$ and $B\sous{trans}$ calculated through spectropolarimetric inversions for the pixel marked with the yellow cross in Fig.~\ref{fig:spmag}. Red lines represent measurements and results for the raster taken between 09:15 and 09:25\,UT while the black lines represent the same pixel of the raster taken 10\,min earlier.}
    {\label{fig:profs}}
\end{figure}

Figure 3 shows the evolution of the total unsigned, net, positive and negative magnetic flux densities, as measured by MDI. For the calculations, only pixels with magnetic field strength higher than 40\,G are considered. The magnetic flux content of the region is in the range of $10^{18}$\,Mx, typical for the quiet Sun and at the lower end for ephemeral regions \citep{Hagenaar01}. The first signs of new magnetic flux appeared around 08:15\,UT, but during roughly the first 20 minutes the increase in total unsigned flux was subtle ($2.2\pm1.6\times10^{14}$\,Mx~s\sur{-1}) and the negative polarity was dominant. Both polarities then increased in concert until 09:00\,UT, thus contributing to an order of magnitude higher increase rate ($2.18\pm0.19\times10^{15}$\,Mx~s\sur{-1}). After that, the positive polarity dominated and the total unsigned magnetic flux continued to increase at a lower rate ($5.1\pm1.8\times10^{14}$\,Mx~s\sur{-1}) until about 09:30\,UT, whereas the negative polarity was fairly constant throughout. Eventually, new magnetic flux stopped accumulating at 09:30\,UT, when both the total unsigned, positive and net flux reached a maximum value. After that, the magnetic flux underwent a slow decay ($-1.15\pm0.30\times10^{14}$\,Mx~s\sur{-1}). The overall evolution profile resembles the evolution of active regions \citep{verma12}, although for the EFR of our study this evolution occurs on timescales of hours instead of days. The rates of fast increase and slow decay are comparable to the ones derived for the evolution of a small-scale bipole studied by \citet{guglielmino12}, indicating a common evolution and decay process.

\begin{figure*}[htp]
\centering
  \includegraphics[width=18.5cm]{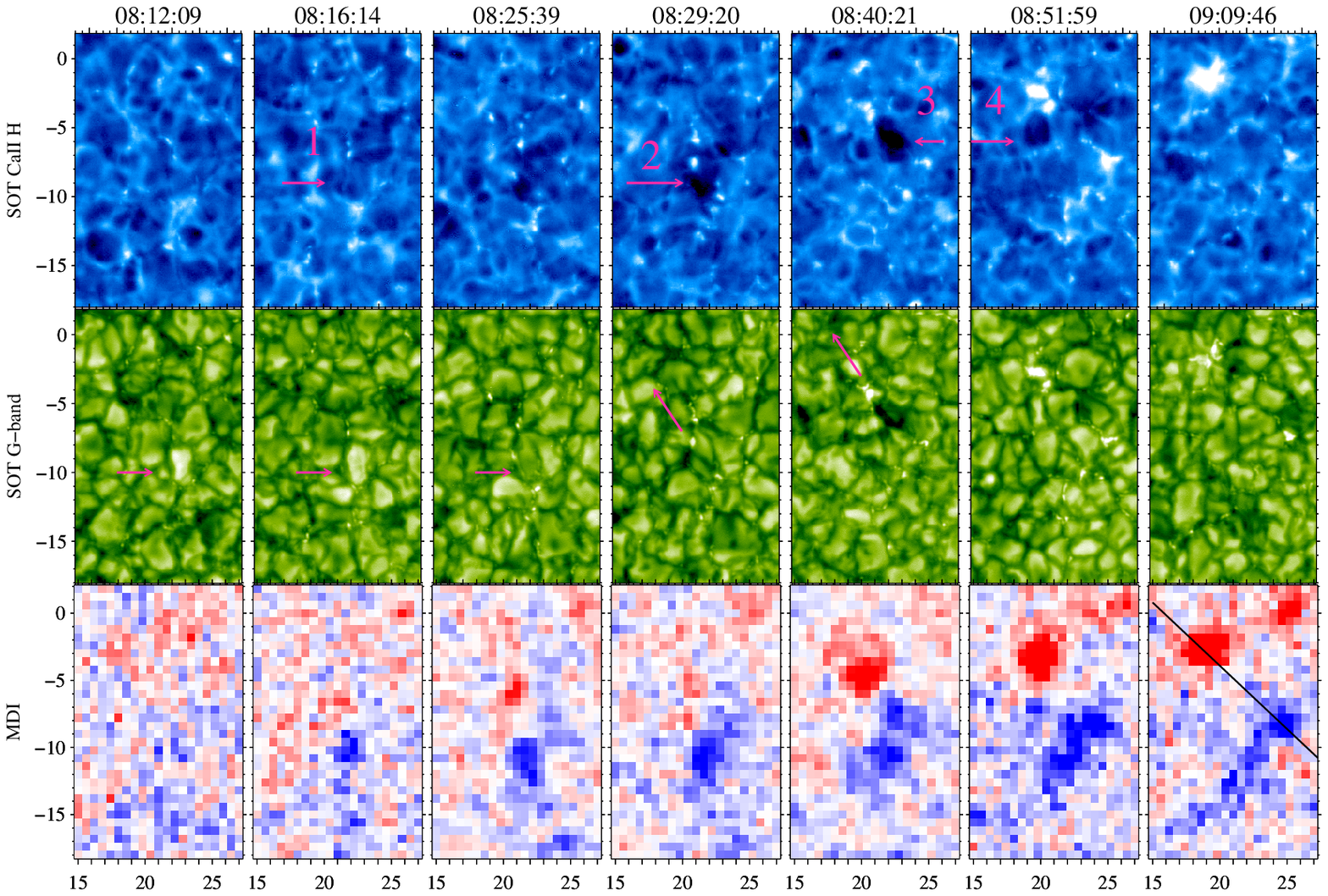}
  \caption[]
    {Overview of the evolution of the EFR (green rectangle in Fig.~\ref{fig:mdi_snaps}) at the lower atmosphere, in Hinode/SOT \caiih\ (top row), G-band (middle row), and MDI (bottom row). The line in the last panel of the bottom row marks the area where the space-time slices of Fig.~\ref{fig:slices} were taken. Arrows indicate specific emergence events and features of interest (see main text). The magnetic field strength is scaled between $\pm$40\,G.}
    {\label{fig:cagb}}
\end{figure*}

This evolution indicates the existence of discrete phases in the lifetime of an EFR. The slow initial pile-up of magnetic flux is in agreement with simulation experiments showing that new granular-scale magnetic flux stops for a few tens of minutes at the base of the photosphere before emerging \citep{cheung07}. The period of fast increase in magnetic flux then follows, which coincides with that of the fast lateral expansion of the bipolar region. After the bipole reached its maximum separation at the photosphere, the magnetic flux started accumulating slowly and then it was followed by a slow decay phase. The observed flux imbalance can be misleading since the negative polarity is apparently comprised of small-scale magnetic elements and probably most of them evade detection by MDI. This observational limitation may also affect the initial stages of the emergence, when only very small-scale magnetic elements appear in the photosphere (as shown in Section~\ref{sec:phot}). 

\subsection{Spectropolarimetry of the emerging region}
\label{sec:spec}

Although the opposite-polarity patches reached their largest separation at around ~09:00\,UT, the region continued to evolve, not only in the upper layers but also in the photosphere. In Fig.~\ref{fig:mag_evo}, we show that magnetic flux continued to accumulate for roughly another 30\,min, before the slow decay phase. During those 30\,min, the Hinode/SP recorded the Stokes spectra of the \fei\ 6301 and 6302\,{\AA} line pair, performing two scans of the region separated by 10\,min.

During the acquisition of the first scan, between 09:05\,UT and 09:15\,UT, the bipolar region had already reached the size of a mesogranular-scale ephemeral region (left panels in Fig.~\ref{fig:spmag}). The $B\sous{LOS}$ maps of the first SP scan show mixed small-scale polarities in the region between the two main opposite-polarity concentrations (top row in the left panels of Fig.~\ref{fig:spmag}). These are analogous to the serpentine magnetic fields in the photosphere of emerging active regions \citep{pariat04}. The tomography of the region shows that the localized magnetic patches at $\log\,\tau = -1$ expand with height (decreasing optical depth), as per the pressure balance between the magnetic flux-tubes and the ambient atmosphere. The overall morphology of the region is indicative of the abundant low-lying magnetic flux tubes and the horizontal magnetic fields at the photosphere \citep{martinez_gonzalez07,lites08}, since the maps of the transverse magnetic flux (middle row in Fig.~\ref{fig:spmag}) show increased amounts of transverse magnetic flux between the two main polarities. The velocity field at the emerging region shows predominant upflows between the two polarities ($\log\,\tau = -1,-2$). The main polarities as well as some of the mixed polarities within the EFR are situated in regions of strong downflows at $\log\,\tau = 0$ and $-1$.

Ten minutes later, during the second scan, the situation was different (top row, in the right panels of Fig.~\ref{fig:spmag}). Although opposite polarities were still scattered in and around the emerging region, the area between the two main polarities was somewhat ``clearer',' indicating that most of the flux system emerged. In addition, the amount of transverse magnetic field was lower, the predominant photospheric upflows at the photosphere were no longer present, and a more regular ``undisturbed'' photospheric pattern was established.

\begin{figure}[htp]
\centering
  \includegraphics[width=8cm]{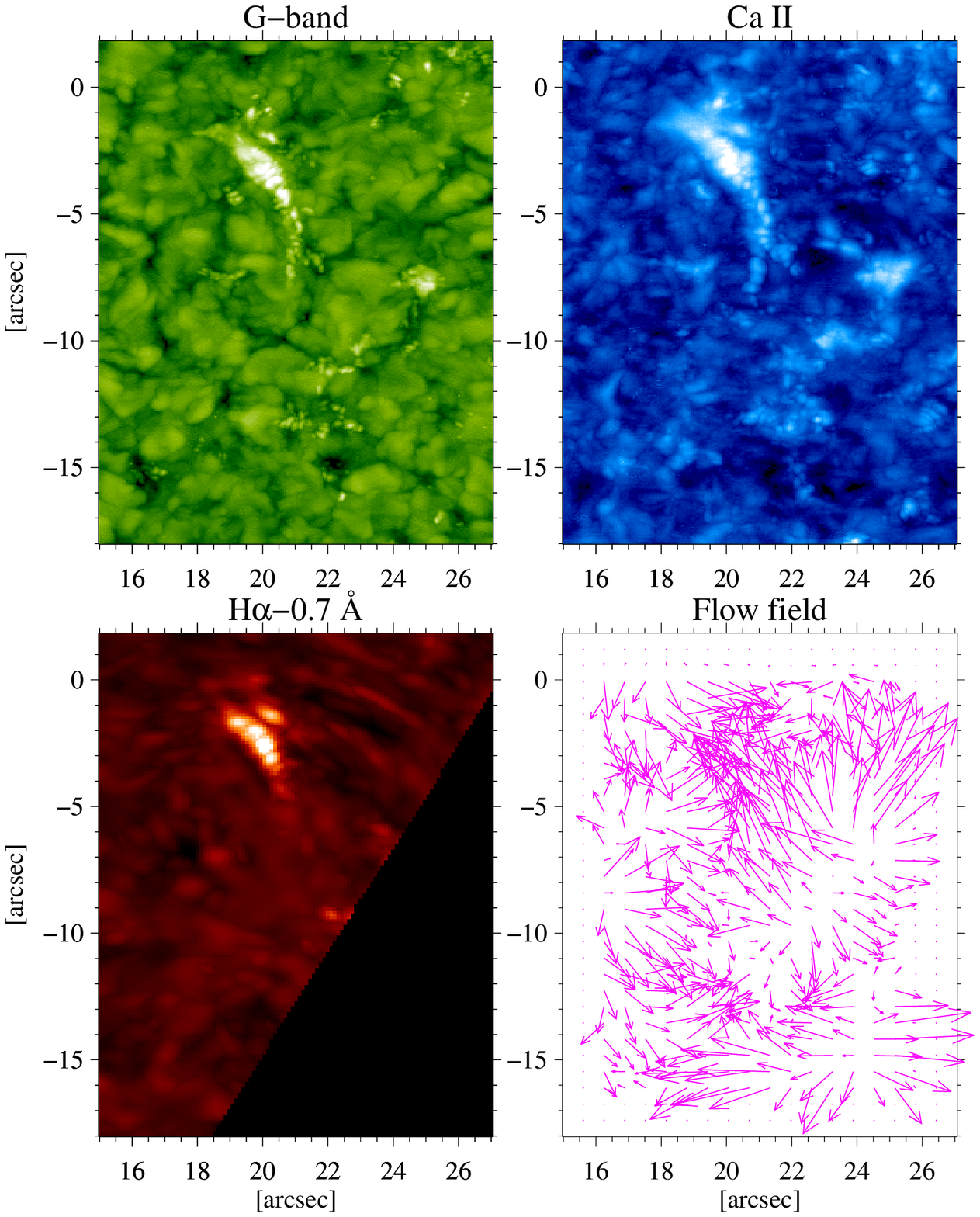}
  \caption[]
    {Maps of temporal maximum values of G-band, \caiih,\ and H$\alpha$--0.7\,{\AA}. Plotted on the bottom right panel is the average flow field, during the first hour of observations, derived through LCT from the G-band observations.}
    {\label{fig:minmax}}
\end{figure}

A feature of interest is seen in both scans at around (23\arcsec,-10\arcsec). This region shows a positive-polarity patch attached to a part of the negative polarity. During the second raster the positive polarity was stronger and more extended and was accompanied by strong upper photospheric upflows. In Fig.~\ref{fig:profs}, we provide the Stokes \textit{V} profile, the stratification of the magnetic field and the LOS velocity within a pixel marked with a yellow cross in Fig.~\ref{fig:spmag}. Several other pixels of the area exhibited similar behavior. In the upper left panel of Fig.~\ref{fig:profs}, the Stokes \textit{V} profiles have an irregular three-lobed shape, indicative of unresolved structure or strong vertical gradients of velocity and field inclination \citep{bellotrubio19}. The $\tau$ profiles of $B\sous{LOS}$ and $V\sous{LOS}$ produced by the inversion exhibit a sign reversal at about $\log\,\tau = -1$. This irregular pattern did not appear 10\,min earlier, indicating that during the evolution of this feature, positive flux was piling up in the vicinity of the negative footpoint leading to a possible interaction between the two polarities.

\subsection{Flux emergence at the photosphere and lower chromosphere}
\label{sec:phot}

In Fig.~\ref{fig:cagb}, we plot snapshots of the evolution of the emerging regions in \caiih\ and G-band at certain times of interest, along with the corresponding frames of the MDI $B$\sous{LOS} maps, for comparison. The G-band (middle row) shows the typical granulation pattern of the photosphere with several very small-scale bright points located at the intergranular lanes. The \caiih\ wide-band filter of SOT shows the reversed granulation pattern observed at the upper photosphere/lower chromosphere \citep{rutten04b}, on top of which bright points are distinctly visible. 

At 08:12\,UT, a small,  bright, slightly elongated granule is observed at ~(22$\arcsec$,-10$\arcsec$), which became more elongated during the following minutes (next frame) and bright points appeared at its two edges, visible both in G-band and Ca\,{\sc ii}\,{\sc H}. The bright point at the upper edge had no clear counterpart in MDI, probably because its magnetic flux density was below the detection limit. However, the bright point at the lower edge of the granule corresponded to a clearly discernible negative-polarity feature.  
The granule then expanded, its inner part darkened indicating the onset of a collapsing process, and in few minutes (around 08:25\,UT) the granule was already fragmented. A bright lane appeared along its left side (see arrow) and a series of bright points pervaded the granule after its fragmentation. The positive polarity started to strengthen and from this point on, an expanding magnetic bipolar region appeared in the MDI maps (compare with Fig.~\ref{fig:mdi_snaps}). 

During the emergence process, a series of dark patches was detected at the upper photosphere/lower chromosphere in \caiih\ (Fig.~\ref{fig:cagb}). These dark regions, marked with arrows in the \caiih\ panels at 08:29, 08:40, and 08:51\,UT, appear between (groups of) bright points and their length and width increased within the following minutes. The bright points located around these patches were moving in opposite directions along the main axes of the patches, as indicated by the arrows in G-band at 08:29 and 08:40\,UT. At least four such events, signified by dark chromospheric patches, were seen in Ca\,{\sc ii}\,{\sc H} (marked with numbers 1--4). The long axis of these patches was parallel to the orientation of the emerging loops \citep{tortosa_andreu09}. Following the second of these events (``2'' at 08:29\,UT), the two polarities became gradually more separated, as seen in the MDI panels that follow, indicating that this is the main part of the emerging structure corresponding to the fast increase in magnetic flux (see Section~\ref{sec:mdi}). Although one compact positive polarity was eventually formed, the negative polarity was more scattered, as if each smaller event (or bundles thereof) produced a different negative-polarity footpoint. Towards the end of these SOT observations, which lasted until 09:09\,UT, another positive-polarity patch appeared next to the main one, likely also formed by the coalescence of smaller bright points, advected there by the plasma motions of the granulation. The whole region extended to mesogranular scale but numerous bright points, which signified fine-scale magnetic concentrations, were found around and within the region indicating further evolution on smaller scales. This was illustrated in Section~\ref{sec:spec}.

One way to visualize the overall evolution of the region and the motions of the bright points in these layers is through maps of maximum intensity. These maps were constructed by calculating  the temporal maximum for every pixel throughout the entire time-series. Since bright points are brighter than the background in G-band, \caiih,\ and in the blue wing of the H$\alpha$ \citep{leenaarts06}, the trajectories of bright points in these wavelengths will appear as bright streaks within the FOV. This is indeed the case for the main positive polarity patch (Fig.~\ref{fig:minmax}). The emerging bright points followed a curved path until reaching the region where they accumulated to form the positive-polarity patch, as seen in all three maps of Fig.~\ref{fig:minmax}. The streaks are thinner in the G-band map, whereas in Ca\,{\sc ii}\,{\sc H} they appear wider and hazier. The latter map also shows more features, some related to the negative-polarity footpoints and some related to the second positive-polarity patch that developed at the upper right corner of the FOV. Some of these brightenings have been attributed to the interaction between the emerging and pre-existing magnetic field of the lower chromosphere \citep{guglielmino08}. The maximum intensity map constructed for the blue wing of H$\alpha$ also shows the path of the positive-polarity magnetic elements, although shorter because the H$\alpha$ observations started about 30 minutes later. It also shows persistent, overlying, dark, elongated features, which are absorption structures related to the chromospheric response to the emergence of new magnetic flux and give the impression of an absorption front in the wake of the moving polarities of the EFR. The chromospheric response is discussed in more detail in Section~\ref{sec:chrom}.

\begin{figure}
\centering
  \includegraphics[width=9cm]{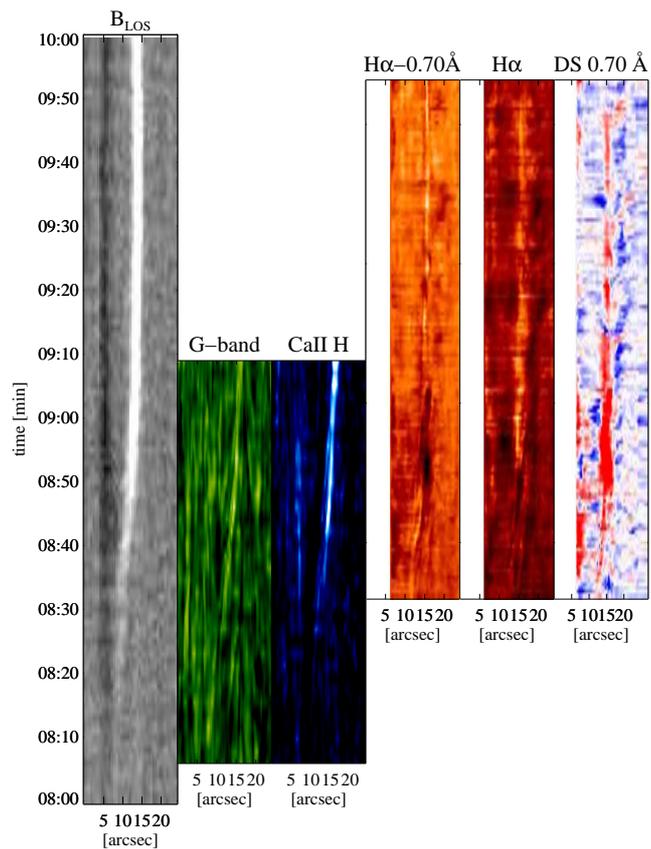}
  \caption[]
    {From left to right: $x$-$t$ slices of MDI LOS magnetic field, G-band, Ca\,{\sc ii}\,{\sc H}, H$\alpha - 0.7\,{\AA}$, H$\alpha$ line core, and H$\alpha$ Doppler signal at $\pm0.7$\,{\AA} from the line center. The slices are placed along a common vertical time axis.}
    {\label{fig:slices}}
\end{figure}

In Fig.~\ref{fig:minmax} we also plot the time-averaged photospheric horizontal flow field. The flow field shows that the emerging region resides on top of a mesogranular-scale diverging flow. The flows are directed towards regions where magnetic flux is concentrated. The trajectories of the bright points follow these flows. As already implied by the examination of the snapshots in Fig.~\ref{fig:cagb}, the two polarities did not move with the same velocity. The velocities reached up to 1.7\,km\,s\sur{-1} (median velocity is 1\,km\,s\sur{-1}) for the part of the mesogranule where positive-polarity magnetic fields accumulate, and up to 0.8\,km\,s\sur{-1} (median velocity is 0.4\,km\,s\sur{-1}) for the negative polarity. These values are much lower than the ones derived for small-scale emergence events in active regions \citep[see e.g., ][]{guglielmino10}, perhaps because these were part of a weaker magnetic flux system. It should be noted that the choice of sampling window, the spatial resolution, and the cadence of the observations affect the LCT velocities. However, our derived horizontal speeds are also supported by the horizontal separation speed estimated by the $x$-$t$ slices discussed in the following section.

\begin{figure}
\centering
  \includegraphics[width=9cm]{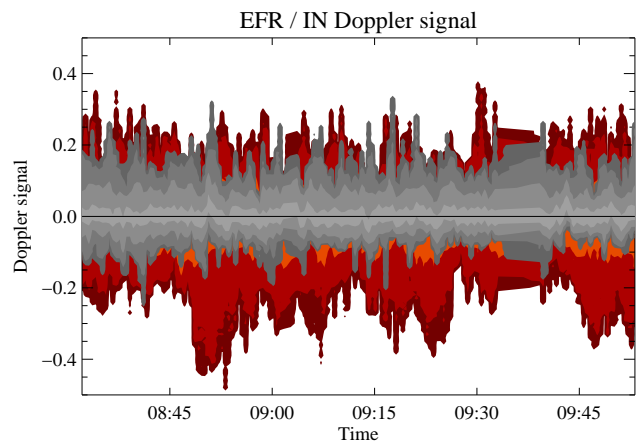}
  \caption[]
    {Temporal evolution of the DS distribution within the EFR (red) in comparison to the temporal evolution of an internetwork region (gray).}
    {\label{fig:ds_evo}}
\end{figure}

In summary, at the photosphere, flux emergence happened repeatedly, building up the flux of the newly formed region. As a result of this episodic emergence, both polarities had significant substructure: the more compact positive polarity comprised finer bright points while the negative polarity was scattered in separated footpoints. Even though discrete emergence events were detected through photospheric imaging, the magnetic flux was accumulating in a more continuous fashion. Events that are below the detection limit of current observations likely also contributed to this build-up of magnetic flux. These bundles of flux would be too weak to have exerted an observable effect at the spatial resolution of SOT on the photospheric flow field and could have been advected passively by granulation.   

\begin{figure*}[htp]
\centering
  \includegraphics[width=18cm]{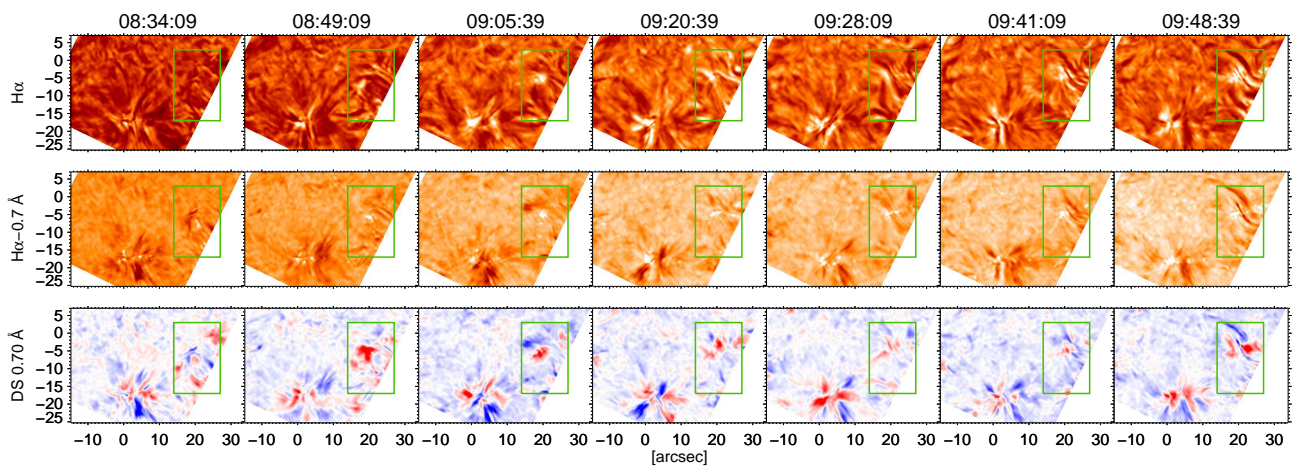}
  \caption[]
    {Chromospheric evolution of the EFR. From top to bottom, running average images over 2.5\,min in H$\alpha$ core (top row) H$\alpha$--0.7\,{\AA} (middle row),  and H$\alpha \pm$0.70\,{\AA} DS (bottom row). The green rectangle marks the region of the EFR and corresponds to the FOV of Fig.~\ref{fig:cagb}. In the DS maps, blue (red) indicates upflows (downflows).}
    {\label{fig:ha}}
\end{figure*}

\subsection{Chromospheric response and subsequent evolution}
\label{sec:chrom}
Figure~\ref{fig:slices} offers a consolidated view of the evolution of the EFR region in the photosphere and the chromosphere via $x$-$t$ slices taken along the main axis of the bipolar feature (black line in Fig.~\ref{fig:cagb}). From 08:30 to 09:00\,UT, the gradually increasing separation distance leads to an estimated horizontal expansion speed of about $\sim$2\,kms\sur{-1}, which is comparable to the velocities derived with LCT (see Fig.~\ref{fig:minmax}). The dark patches between the two polarities for the main emergence event are visible in G-band and Ca\,{\sc ii}\,{\sc H}. In addition, the fine structure of the positive-polarity cluster is clearly shown, the trace of which appears in the $x$-$t$ slices as two braided bright streaks, mostly evident in Ca\,{\sc ii}\,{\sc H}. This feature is also evident in the H$\alpha$ line-wing intensity at $-0.7$\,{\AA}. The positive-polarity bright point appeared in the core of H$\alpha$ several minutes later, at about 08:50\,UT, indicating a time-lag between appearance in the upper photosphere and appearance in the upper chromosphere. Until that time, the positive footpoint was covered by redshifted absorption features. However, there is an effect in the core of H$\alpha$ from the beginning of the H$\alpha$ observations, indicating an almost immediate interaction with the ambient chromosphere as the new flux rises.  

During the expansion phase (Fig.~\ref{fig:slices}, between 08:30 and 09:00\,UT), increased absorption in \ha\ was located ahead of the dominant (positive) polarity, which was mostly blueshifted. In actual filtergrams (see the discussion of Fig.~\ref{fig:ha} later in this section), these absorption features appeared as a front ahead of the emerging structure, as the positive-polarity patch  moved horizontally towards the northeast direction, and made its way into the ambient chromospheric magnetic field. Towards the end of the expansion phase, downflows over the footpoint became more extended and prominent, the footpoint appeared in the H$\alpha$ core and the blueshifted events ahead of the structure became more elongated and more intense, particularly after 09:00\,UT. This is when the EFR has fully evolved and reached maximum chromospheric height. The flux bundles between the two polarities showed a highly variable velocity pattern. Although  strong, mostly blueshifted redshifts also appeared either along the entire loops or in parts of them, indicating instances of flows along the loops (e.g., Fig.~\ref{fig:slices} around 09:00\,UT). Thus, the chromospheric morphology of the EFR is similar to an arch-filament system \citep[AFS, ][]{bruzek67,murabito17,gonzalez_manrique18}; this chromospheric configuration is associated with emerging flux from the smallest granular-scale \citep{centeno17} up to active regions. On top of this characteristic pattern, we also observed signatures of chromospheric ejecta (e.g.,  at around 09:10\,UT in Fig.~\ref{fig:ha}, discussed later in this section).

We can illustrate how the new magnetic flux affects the chromospheric dynamics even before reaching its full extent by plotting the temporal evolution of the DS distribution of the region against that of a ``quiet'' (i.e., with no persistent chromospheric structures) internetwork region (Fig.~\ref{fig:ds_evo}). Acknowledging that we cannot produce velocity values from the H$\alpha$ filtergrams due to the sparse sampling (only five points) of the \ha\ profile, this figure also puts the DS values in context. The internetwork region exhibits slightly asymmetric distribution in favor of upward motions (positive DS), whereas the DS distribution of the EFR is broader and overall more symmetric, exhibiting more extended and frequent upflow excursions, and, at times, intense and persistent downflows. The most prominent of these downflow excursions started a few minutes after 08:45\,UT and can also be seen in the $x$-$t$ slices in Fig.~\ref{fig:slices}. Given that the cadence of our H$\alpha$ observations is 30\,s, we conclude that the chromosphere above this structure was highly dynamic, evolving within tens of seconds. This kind of evolution can be attributed to the interaction between the newly emerging magnetic field and the presumably weak, ambient magnetic field of the quiet Sun. Since the H$\alpha$ observations started only a few minutes after the emergence, this interaction did not start when the bipolar structure reached its maximum height but at least as soon as magnetic flux started to accumulate in the region (i.e., after 08:30\,UT; see Fig.~\ref{fig:mag_evo}) and before the fast expansion of the region. 

\begin{figure*}[htp]
\centering      
  \includegraphics[width=18cm]{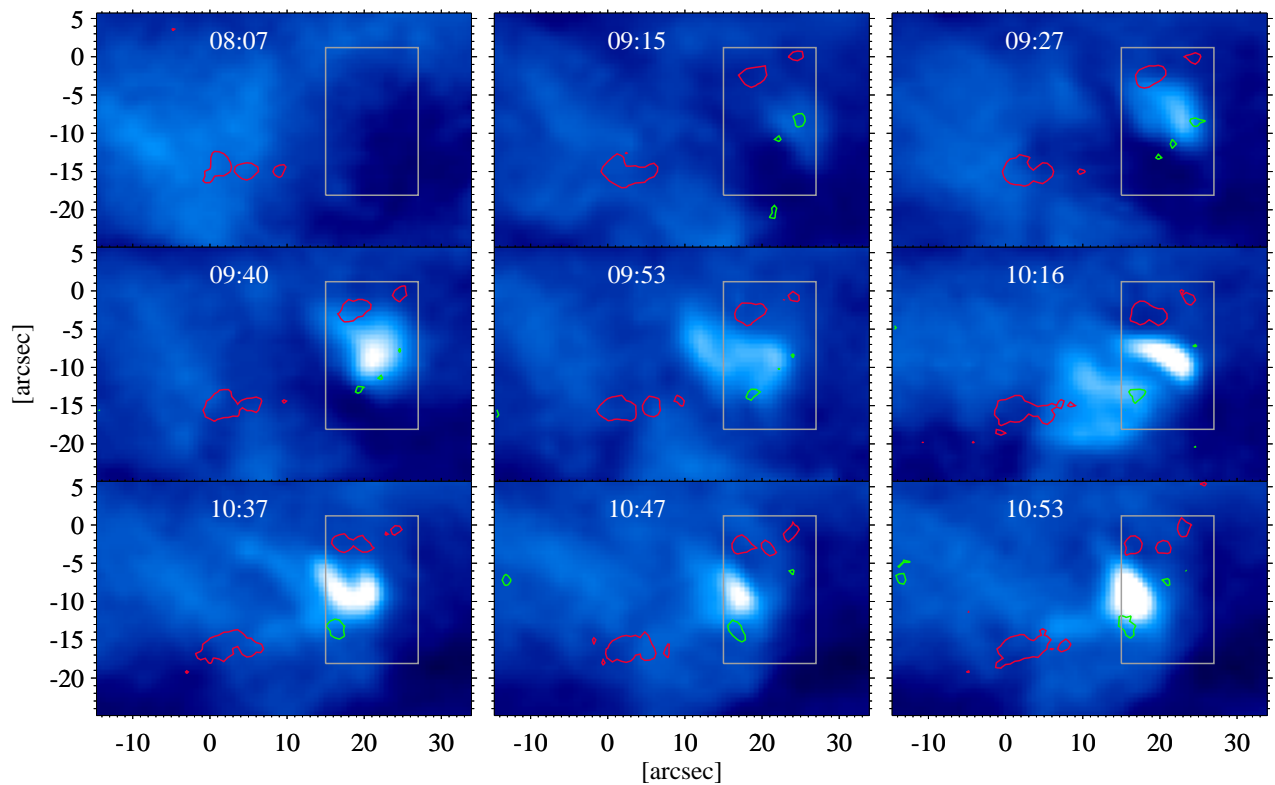}
  \caption[]
    {XRT frames showing the temporal evolution of the coronal bright point. The EFR under study is located within the gray rectangle. Contours mark the $\pm40$\,G levels of the temporally closest LOS magnetic field, as measured by MDI.}
    {\label{fig:xrt}}
\end{figure*}

A more detailed account of the chromospheric evolution of the region and its interaction with the initially undisturbed chromosphere can be acquired through the snapshots plotted in Fig.~\ref{fig:ha}. We plot a larger FOV than in Fig.~\ref{fig:cagb}--\ref{fig:spmag} to include the already formed network boundary to the left of the region, but we mark the FOV of those figures with a green rectangle. As already mentioned, the negative polarity is located outside the FOV of the H$\alpha$ images, but we can monitor the evolution of the positive polarity and some parts of the negative polarity, which gradually approach the already formed network to the left. Initially, the positive-polarity and its horizontal motion towards the northeast direction was visible only at the wing of the H$\alpha$. During this apparent motion, it pushed through the ambient chromosphere, producing absorption features in the wake of its expansion and ejecta at various instances. The blue-shifted absorption front can be seen to surround the (red-shifted) bright point in the first column of Fig.~\ref{fig:ha}. On average, strong downflows were present over the footpoints (see e.g., Fig.~\ref{fig:ha} at 08:49\,UT) while the central parts of the rising loops were mostly blueshifted. This overall upward motion was variable and was often interrupted by red-shifts. 
Around the footpoints and the loops, jet-like blueshifted structures appeared (e.g., at 09:05 and 09:20\,UT). The complicated and highly variable velocity field established at the chromosphere is also more visible in the panels of the bottom row of Fig.~\ref{fig:ha}. 

Another interesting finding is the gradual integration of the new region into the already existing chromosphere. We can infer this from the evolution of the absorption features between the EFR and the network region shown in Fig.~\ref{fig:ha}. Initially, fibrils were more or less radially distributed around the network, but as time progressed more absorption features emanating from its right side were directed towards the negative-polarity part of the EFR. These loop-like structures were sometimes seen to connect the two regions. \citet{kontogiannis18} demonstrated that some of these low-lying loops were heated to transition region temperatures. 

Around 09:21\,UT, a blueshifted, elongated absorption feature appeared over the positive-polarity part of the EFR. A similar structure but more strongly blueshifted and elongated appeared a few minutes later (09:28\,UT) and was visible until the end of the H$\alpha$ observations at 10:00\,UT. Another upward moving feature appeared to pervade the EFR (Fig.~\ref{fig:ha} at 09:48). These chromospheric jet-like features, often called surges, are presumably the result of reconnection between the newly emerged magnetic structure and the ambient coronal magnetic field, while the former gradually peels-off and is assimilated into the already existing chromospheric environment, as shown in recent MHD simulations \citep{nobrega16}. Surges can appear in different stages of the EFR evolution and their orientation is strongly determined by the ambient magnetic field \citep{mactaggart15}. As discussed in the following section, this intense activity in the chromosphere of the EFR coincides with the impulsive brightenings observed in the corona.

\subsection{Coronal bright-point formation and EUV spectroscopy}

\begin{figure}[htp]
\centering
\includegraphics[width=9cm]{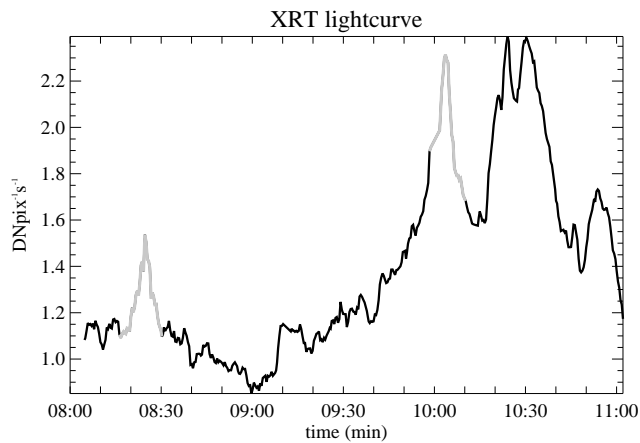}
  \caption[]
    {XRT ``C-Poly'' light curve, taken as the background-subtracted average of the pixels inside the gray rectangle in Fig.~\ref{fig:xrt}. The peaks in gray at around 08:20 and 10:00\,UT are caused by intense spikes in the images.}
    {\label{fig:xrt_lc}}
\end{figure}

The upper atmospheric response to the emergence was monitored continuously by the XRT and intermittently by the EIS. In Fig.~\ref{fig:xrt}, we plot snapshots of the evolution of the brightening observed in soft X-rays during the three hours of the coordinated observations. For comparison, we overplotted contours of the MDI magnetic field strength ($\pm40$\,G), which outline the positive and negative magnetic polarities of the EFR. Figure~\ref{fig:xrt_lc} contains the corresponding light curve calculated for the region. Until about 09:00\,UT, the region was darker than the neighboring network region and its total emission was slightly decreasing. This could be attributed to the new region pushing aside the ambient corona. The new region contained cool plasma and therefore produced this slight decrease. At this region, \citet{kontogiannis18} showed that during the same period there was a gradual increase of emission in the upper chromospheric and transition region spectral lines, but with no coronal counterpart. After 09:00\,UT, an abrupt but small increase was followed by a continuous rise of emission with increasing rate which lasted until 10:30\,UT. The brightening first appeared at the rightmost negative footpoint after 09:20\,UT, and was then expanding and changing shape, covering most of the newly emerged region and part of the neighboring network. After 10:10\,UT, the upper part of the structure brightened abruptly reaching its peak emission 10\,--\,20\,min later. After that, the coronal bright point started to fade, exhibiting a secondary peak at 11:00\,UT. Concluding the discussion on the coronal morphology of the region, we note that the brightening was also visible in the 171\,{\AA} channel in one image acquired by TRACE at 09:37\,UT (not shown).

\begin{figure*}[htp]
\centering
  \includegraphics[width=19cm]{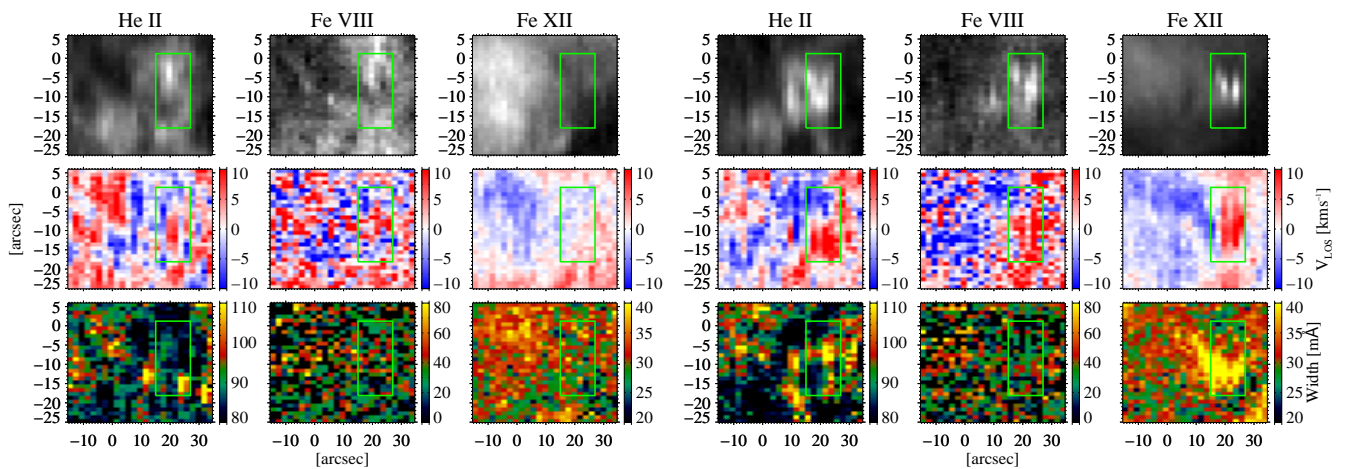} 
  \caption[]
    {From left to right: Spectral-line parameters for the three spectral lines, \heii, \feviii, and \fexii\ of the two EIS scans, the scan between 09:01 and 09:20\,UT, and the scan between 10:25\,--\,10:44\,UT (left and right panels, respectively). Rows from top to bottom contain intensity, LOS velocity, and spectral line width.}
    {\label{fig:eis_int}}
\end{figure*}

The EIS performed two raster scans of the region between 09:01\,and\,09:20\,UT and between 10:25\,and 10:44\,UT, that is, right before the first brightening and during the second one, respectively. In Fig.~\ref{fig:eis_int}, we plot three representative EIS spectral lines of the upper chromosphere, transition region, and corona (\heii, \feviii\, and {Fe~{\sc xii}}, respectively). 
During the first scan (left panels in Fig.~\ref{fig:eis_int}) we detected no considerable coronal emission, while emission in the cooler spectral lines indicated that the structure already started filling with plasma at transition region temperatures. Some of this EUV emission may be also associated with chromospheric jet-like structures that protrude from the newly emerged structure \citep{kontogiannis18}. The velocity maps show predominantly upward velocities (blueshifts). The thermal width is noisier because the thermal width is the third moment of the intensity and the region is quiet. However, patches of higher width around the region in the \heii\ map may result from chromospheric ejecta around the region. During the second scan (right panels in Fig.~\ref{fig:eis_int}), the emission in all spectral lines was enhanced, more extended, and appeared like a double-loop structure. The different spatial extent of the emission in the three spectral lines indicates a temperature stratification within the region, with the coronal emission stemming from the loop tops and the lower emission stemming from the more extended lower parts. All spectral profiles exhibit pronounced redshift and increased thermal widths, especially the \fexii\ line which exhibits an extended high-thermal-width feature over the region. The redshift detected in all three spectral lines (although a common finding for TR temperatures) may be an indication that the EFR slowly subsided, which is also in agreement with the approaching footpoints, as shown in Fig.~\ref{fig:mdi_snaps}.

\begin{figure}[htp]
\centering
  \includegraphics[width=9cm]{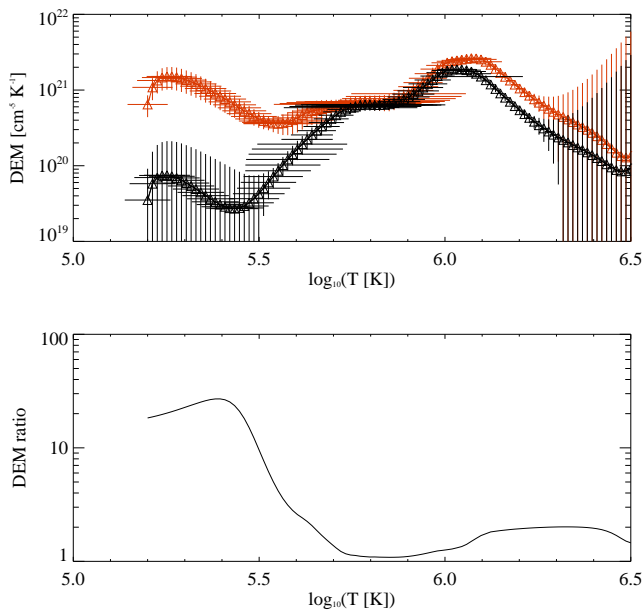} 
 \caption[]
    {Top: DEM analysis for the EFR region during the first (black line) and the second (red line) raster scans. Bottom: Ratio of the two DEM curves.}
    {\label{fig:dem}}
\end{figure}

For the DEM analysis we summed up the EIS pixels over the emerging loop structure visible in the \fexii 192.12\,\AA\ line and computed averaged intensities for all DEM spectral lines of Table~\ref{Table:t2}. 
The Gaussian fit on the average profiles was successful except for the case of the \mgv\ line, which was very weak and noisy.
The ratio of the \fexi\ 192.63/192.80\,
\AA\ lines yielded an electron density of $10^{10}$~cm$^{-3}$. We used this density value and adopted a photospheric abundance \citep{asplund09} and the ionization equilibrium from the Chianti v8 atomic physics database \citep{chianti} to compute the contribution functions of the selected spectral lines. The DEM function was calculated using the method of \citet{hannah12} and the DEM curves of the same position in the two rasters were determined for comparison (Fig.~\ref{fig:dem}). 

The DEM curves are well defined in the temperature range between $\log\,T\,=\,5.2$ and $\log\,T\,=\,6.3$, while for higher temperatures the error bars are high due to the lack of emission. During the brightening, the DEM is higher overall throughout the entire temperature range, exhibiting an increase by a factor of 20 for temperatures lower than $\log\,T\,=\,5.4$ and by a factor of 2 in the range $6.0\,<\,\log\,T\,<\,6.5$. Thus, the plasma heating, although considerable in coronal temperatures, is higher at transition-region temperatures. Considering that the region was scanned between 10:25 and 10:44\,UT, we conclude that the DEM curve of the second raster corresponds to the cooling phase of the brightening (see Fig.~\ref{fig:xrt_lc}), hence the lower increase of the emission measure in coronal temperature and its twenty-fold increase at lower temperatures.

\section{Discussion and conclusions}

We studied the evolution of a new magnetic flux system in the quiet Sun, from its emergence at the photosphere to its appearance at the corona. Our multi-wavelength observations allowed us to monitor different temperature regimes, illustrating the series of events that followed the appearance of new magnetic field structures in the solar atmosphere.

The size of the emerging region is at the low end of the magnetic flux magnitudes for ephemeral regions. Its evolution exhibits distinct phases, namely a slow flux accumulation during the first few minutes, a fast increase during the subsequent 30\,min, and a slower increase of equal duration followed by a slow decay. The magnetic flux rates measured during each phase are in general agreement with previous measurements and exhibit analogies with active regions. 

The fast increase phase is characterized by consecutive small-scale events of granular scale observed at the photosphere, whose coalescence produced a largely bipolar magnetic configuration. During each of these events, we witnessed the typical features of emergence at the photosphere, namely elongated granules and dark patches at the upper photosphere and lower chromosphere. Photospheric spectropolarimetry during the last phase of flux accumulation revealed a pattern resembling that of the serpentine-type magnetic fields observed in active regions.  These observations support the emergence of a distorted flux tube, whose fragments appeared and merged at the photosphere.  

The chromospheric dynamic evolution appeared very early, slightly preceding the fast increase phase; it comprised fast, intense upflow events over and around the EFR and long-duration, intense downflows above the footpoints. These events were more intense and dynamic compared to the typical internetwork regions. As the region expanded and the footpoints moved horizontally, absorption fronts were forming ahead of the region where the chromospheric material exhibited upward motions. The positive-polarity footpoint itself was well established at the chromosphere roughly 30\,min after emergence. The newly emerged region was gradually consolidated into the ambient chromosphere by forming connections with the neighboring, pre-existing network. These connections had the appearance of loop-like chromospheric structures which were heated to transition-region temperatures, as shown in a previous study by \citet{kontogiannis18}. 

Soft X-ray imaging showed a weak dimming during the first 30\,minutes of the emergence, which could be an indication that the new region was pushing aside the ambient corona. This dimming was followed by a gradual increase of coronal emission in conjunction with more impulsive brightening events. Comparison with photospheric spectropolarimetry showed that flux cancelation at the negative-polarity footpoint contributed to the gradual increase of coronal emission. The position and the timing of the coronal brightening support this scenario. Co-temporal \ha imaging indicates the formation of surge- or macrospicule-like structures during the impulsive coronal brightenings. The DEM analysis of the region captured the cooling phase of the brightening, showing only a small increase in coronal temperatures and a significant increase of emitting plasma below $\log\,T\,<\,5.7$, combined with dominant receding motions, throughout all temperatures.  

Our observations did not capture the complete decay of the region. However, towards the end of the observations, MDI magnetograms showed that the opposite-polarity patches eventually fragmented; parts of them approached, while others interacted with neighboring structures. It is likely that part of the region eventually diminished and subsided. This type of evolution was also observed in larger EFRs by \citet{verma16} and \citet{gonzalez_manrique17}. 

Several features predicted by recent MHD numerical modeling are confirmed by our study. The absorption fronts, which were detected for the first time, can be attributed to the acceleration of chromospheric material by strong gradients of thermal and magnetic pressure during the horizontal motion of the positive-polarity magnetic footpoint, as demonstrated by \citet{yang18}. This effect appears at the first stage of emergence and can produce cool jets and subsequent shocks at the overlying corona. We surmise that part of the intense activity observed during the expansion of the EFR can also be attributed to this effect, which produces, through shock-heating, the gradual increase in the transition-region emission reported by \citet{kontogiannis18} and also seen here as a smooth increase in the soft X-ray output (Fig.~\ref{fig:xrt_lc}). The \ha surge observed later in the EFR evolution was likely driven by reconnection; the ejected material followed the ambient magnetic field, as demonstrated by \citep{mactaggart15}. Since the EFR is not located in a coronal hole, this weak ambient magnetic field is likely tilted, explaining the elongation and orientation of the chromospheric structure. The hot counterpart of these events contributed to the further brightening of the region in the corona, as demonstrated by \citet{nobrega16}. This interrelation between explosive phenomena in different temperature regimes occurs frequently during magnetic cancelation events within active regions \citep{madjarska09,nobrega16,guglielmino18}, and we demonstrated it also in the quiet Sun.

During its short lifetime, the coronal emission exhibited intense spatial and temporal variability, which can be explained by several factors. First, different types of reconnection during the lifetime of a coronal structure can produce stronger or weaker events \citep[e.g.,][]{zhang12}. Our case indicates that flux cancelation (see Fig.~\ref{fig:spmag}) produced the first brightening. The drift of the negative-polarity footpoints of the EFR towards the positive-polarity network and the resulting interaction may also explain the expansion of the coronal brightening towards the left side (see Fig.~\ref{fig:xrt}). Another cause of the spatial variability could be the absorption by cool overlying material, as seen for example in the case of active region moss \citep{depontieu09}. Elongated chromospheric structures found to pervade the region could also produce this effect (Fig.~\ref{fig:ha}). Similar dimmings can be produced by mini-CMEs \citep{innes09} and/or mini-filaments \citep[see e.g.,][]{panesar17} and recent high-resolution observations revealed that eruptions from coronal bright points are common \citep{mou18}, and their magnetic topologies are complicated enough to support relevant driving mechanisms \citep{galsgaard19}. We cannot conclude if this applies to our case since our EFR has a much shorter lifetime and smaller spatial extent than the coronal bright points usually studied; the high-resolution magnetic field observations that would allow modeling are sparse. 

In conclusion, it appears that even at the smallest scales, the emergence of magnetic flux is followed by intricate interactions that produce diverse energetic phenomena at the most quiet solar conditions. These phenomena are close to the detection limits of current instrumentation but seem to contribute significantly to the heating and mass of the upper solar atmosphere. More observations, especially with the new generation of observing facilities underway, such as the Daniel K.Inouye Solar Telescope \citep[DKIST,][]{Tritschler16} and the European Solar Telescope \citep[EST,][]{Collados10a}, along with realistic modeling based on high-resolution magnetic field observations, should lead to accurate estimates of the contribution of these mechanisms to the atmospheric energy budget.

\begin{acknowledgements}
We would like to thank the anonymous referee for providing useful suggestions and constructive comments, which improved the content of the manuscript. This work was supported by grant DE~787/5-1 of the Deutsche Forschungsgemeinschaft (DFG) and by the European Commission's Horizon 2020 Program under grant agreements 824064 (ESCAPE -- European Science Cluster of Astronomy \& Particle physics ESFRI research infrastructures) and 824135 (SOLARNET -- Integrating High Resolution Solar Physics). KT acknowledges support of this work by the project ``PROTEAS II'' (MIS 5002515), which is implemented under the Action ``Reinforcement of the Research and Innovation Infrastructure'', funded by the Operational Programme ``Competitiveness, Entrepreneurship and Innovation'' (NSRF 2014-2020) and co-financed by Greece and the European Union (European Regional Development Fund). The DOT was operated at the Spanish Observatorio del Roque de los Muchachos of the Instituto de Astrof\'{i}sica de Canarias. The authors thank P. S\"{u}tterlin for the DOT observations and R. Rutten for the data reduction. Hinode is a Japanese mission developed and launched by ISAS/JAXA, collaborating with NAOJ as a domestic partner, and NASA and STFC (UK) as international partners. Scientific operation of the Hinode mission is conducted by the Hinode science team organized at ISAS/JAXA. This team mainly consists of scientists from institutes in the partner countries. Support for the post-launch operation is provided by JAXA and NAOJ (Japan), STFC (U.K.), NASA, ESA, and NSC (Norway). Hinode SOT/SP Inversions were conducted at NCAR under the framework of the Community Spectro-polarimetric Analysis Center (CSAC; http://www.csac.hao.ucar.edu). CHIANTI is a collaborative project including George Mason University, the University of Michigan (USA), and the University of Cambridge (UK).

\end{acknowledgements}

\bibliographystyle{aa} 
\bibliography{references}

\end{document}